\documentstyle[twoside]{article}

\catcode`\@=11 \long\def\@makefntext#1{
\protect\noindent \hbox to 3.2pt {\hskip-.9pt
$^{{\eightrm\@thefnmark}}$\hfil}#1\hfill}       

\def\thefootnote{\fnsymbol{footnote}}
\def\@makefnmark{\hbox to
0pt{$^{\@thefnmark}$\hss}}                     

\def\ps@myheadings{\let\@mkboth\@gobbletwo
\def\@oddhead{\hbox{}
\rightmark\hfil\eightrm\thepage}
\def\@oddfoot{}\def\@evenhead{\eightrm\thepage\hfil
\leftmark\hbox{}}\def\@evenfoot{}
\def\sectionmark##1{}\def\subsectionmark##1{}}

\def\qed{\hbox{${\vcenter{\vbox{ 
height
0.4pt\hbox{\vrule width 0.4pt height 6pt
\kern5pt\vrule width 
0.4pt}\hrule
height 0.4pt}}}$}}

\renewcommand{\thefootnote}{\fnsymbol{footnote}} 

\input{mpla1.sty}

\def\@refcitex[#1]#2{\if@filesw\immediate\write\@auxout  
	{\string\citation{#2}}\fi 
\def\@citea{}\@refcite{\@for\@citeb:=#2\do 
	{\@citea\def\@citea{, }\@ifundefined 
	{b@\@citeb}{{\bf ?}\@warning 
	{Citation `\@citeb' on page \thepage \space undefined}} 
	\hbox{\csname b@\@citeb\endcsname}}}{#1}} 
 \def\@refcite#1#2{{[{#1}]\if@tempswa\typeout     
        {WSPC warning: optional citation argument 
	ignored: `#2'} \fi}} 
 \def\refcite{\@ifnextchar[{\@tempswatrue 
	\@refcitex}{\@tempswafalse\@refcitex[]}}

\def\barr{\begin{array}} 
\def\earr{\end{array}} 
\def\berr{\begin{eqnarray}}
\def\err{\end{eqnarray}} 
\def\be{\begin{equation}} 
\def\ee{\end{equation}}
\def\fr{\frac} \def\no{\nonumber}

\newcommand{\pder}[2]{\frac{\partial{#1}}{\partial{#2}}}
\newcommand{\dd}[1]{\frac{\partial}{\partial {#1}}}

\renewcommand{\l}{\lambda}
\renewcommand{\L}{\Lambda} \renewcommand{\t}{\theta} 

\renewcommand{\d}{\delta} \newcommand{\G}{\Gamma} 
\newcommand{\g}{\gamma}
\newcommand{\del}{\partial}
\renewcommand{\phi}{\varphi}
 \renewcommand{\o}{\omega} 

\begin{document} 
\setlength{\textheight}{7.7truein} 

\runninghead{Motion of a vector particle$\ldots$}
{Motion of a vector 
particle
$\ldots$}

\normalsize\textlineskip \thispagestyle{empty}
\setcounter{page}{1}

\copyrightheading{} 

\vspace*{0.88truein}

\fpage{1} \centerline{\bf MOTION OF A VECTOR PARTICLE
IN A CURVED 
SPACE-TIME.}
\baselineskip=13pt \centerline{\bf I.  LAGRANGIAN
APPROACH.}
\vspace*{0.37truein} \centerline{\footnotesize ZAFAR
TURAKULOV\footnote{Institute of Nuclear Physics,
Ulugbek, Tashkent 
702132,
Uzbekistan.  \mbox{E-mail:zafar@suninp.tashkent.su}}} 
\baselineskip=12pt
\centerline{\footnotesize\it Inter-University Centre
for Astronomy and
Astrophysics,} \baselineskip=10pt
\centerline{\footnotesize\it Post Bag 
4,
Ganeshkhind, Pune 411007, India }
\vspace*{10pt}

\centerline{\footnotesize MARGARITA
SAFONOVA\footnote{E-mail:  
rita@ducos.ernet.in}} \baselineskip=12pt
\centerline{\footnotesize\it Department of Physics and
Astrophysics,
University of Delhi} \baselineskip=10pt
\centerline{\footnotesize\it 
New
Delhi--7, India} \vspace*{0.225truein}

\publisher{(received date)}{(revised date)}
\vspace*{0.21truein}
\abstracts{From a simple Lagrangian the equations of
motion for a
particle
with spin are derived.  The spin is shown to be
conserved on the 
particle's world-line.  In the absence of a spin the equation
coincides with that 
of a
geodesic.  The equations of motion are valid for
massless particles as 
well,
since mass does not enter the equations
explicitly.}{}{}




\setcounter{footnote}{0}
\renewcommand{\thefootnote}{\alph{footnote}}

\vspace*{1pt}\textlineskip \section{Introduction}
\vspace*{-0.5pt}

According to the equivalence principle (EP) of the
general theory of
relativity (GR), motion of structureless test
particles in a 
gravitational
background is determined only by the spacetime
geometry: particle
worldlines are the geodesics of the spacetime. Things
become more 
complicated
for test particles which are not structureless and
carry, for example, 
a
non-vanishing charge or spin. In such cases, the
worldline of a
particle, in general, is no longer a geodesic, but is
modified by
electromagnetic and/or spin-gravity forces (for, e.g., 
\refcite{Pomeranski,Balakin} 
and
references therein). The problem of the motion of {\it
classical} 
spinning 
particle in external fields has occupied scientists
for nearly all of 
the last  
century. Covariant equations of motion for a
relativistic particle in 
an 
electromagnetic field were first written more than 70
years ago by 
Frenkel  
\cite{Frenkel}. For the case of a gravitational field,
Mathisson 
\cite{Mathisson} and Papapetrou \cite{Papapetrou}
found
the energy-momentum and angular momentum propagation
equations for a 
rotating
test body (``pole-dipole particle") according to the
Einstein's GR. 
Tulczyjew
\cite{Tulczyjew}, Beiglboeck \cite{Beiglboeck} and
Madore \cite{Madore} 
developed these into laws of
motion by adding a definition of a centre-of-mass
world line. Later, 
Dixon
\cite{Dixon} generalized these treatments and made
them more rigorous. 
All 
theses works dealt with the motion of extended
rigid bodies or 
tops. 
It is far from obvious whether one can observe in
practice the spin corrections to the equations of
motion of elementary
particles. However, the problem of influence of the
spin on the 
trajectory of
a particle in an external field is not of only
theoretical interest.
Spin-dependent corrections certainly exist in
differential cross 
sections of
scattering processes. It was proposed long ago to
separate charged 
particles
of different polarizations through the spin
interaction with external 
fields
in a storage ring of accelerator \cite{Niinikoski}.
Though this 
proposal is
being discussed rather actively now (see a review
 in Ref. \refcite{Heinemann}) it 
is not 
yet
clear whether it is feasible technically. The EP can
be put to test
in an astrophysical setting, a recent proposal being
based on the 
analysis of 
the
differential time delay between the arrival of left
and right-handed
circularly polarized (LCP and RCP) signals from the
millisecond pulsar 
PSR
1937+214 \cite{Biplab}. However, far few papers exist
on the theoretical foundations of 
possible
deviations of the photon motion from geodesics (see,
e.g., Ref. 
\refcite{Biplab,Prasanna}). Here we report yet another
approach to this
long-standing problem. If an appropriate
Lagrangian density is taken into account, then the
photon equation of 
motion
(modified geodesic equation) can be found from the
Euler-Lagrange 
equations.

\vspace*{1pt}\textlineskip \section{Formalism}
\vspace*{-0.5pt}

The generalized concept exists that classical
particles follow the path 
of
the least spacetime distance between the endpoints,
even when space is 
curved
by gravity. Thus, in (pseudo-) Riemannian spaces the
geodesic equation 
is
found from the variation of an action $S$, identified
with the
parameter $s$ of a curve interpreted as its length.
The same method has 
been
used to find the geodesic equation of light where,
however, one 
technical
problem arises: photon's worldline has no length. One
way to avoid this
difficulty is to consider the motion of a massive
vector particle and, 
if the
obtained equation does not contain mass explicitly,
simply put the mass 
to
zero; with one further condition that the four
velocity of a particle 
be a
null vector. This method has been widely used for
scalar particles and 
is
known to give the equation of geodesic regardless of
mass. In this work 
we
apply this approach, which makes it possible to use
the length 
parameter 
of the action principle.

Usually in order to describe the behavior of a field
in a given 
gravitational
background, one solves the corresponding field
equations for a given 
metric.
If the goal is to describe waves, one can take the
corresponding wave
solution. In a spherically symmetric spacetime the
solutions contain 
factors
expressed in terms of spherical harmonics. This method
works well when 
the
wavelength is comparable to the scales under
consideration. It is not 
so in
the case of light propagating in the vicinity of a
massive object and, 
to
describe the propagation of light as an
electromagnetic field, one 
would have
to employ the spherical harmonics of a very high
order. The 
corresponding
solution would look too complicated and tell little
about the behavior 
of
light.

Many efforts have been spent in the last decades to
work out a simple 
approach
to this problem that would give a satisfactory
approximation to the 
wave as
some curves that could be called ``rays" and, at the
same time, would 
take
into account the polarization of light. In the present
work we try to 
work
out a simple approximation of this type. Our idea
consists of the 
following:
we consider a massive vector field obeying the Proca
equation, 
describing the
propagation of this field, in some restricted domain
of the spacetime. 
The
shape of the domain can be chosen as that of some
world-tube transverse 
to the
wave, with the cross-section comparable to the
wavelength, or, say, not
greater than two orders of magnitude. As this tube is
timelike there 
must
exist a timelike curve $\L$ in its interior, which
specifies a local
coordinate system as the time axis. And if the tube is
not too wide, 
this
coordinate system would cover the entire interior. If
$s$ is the 
proper time
on the curve $\L$, the curve can be chosen in such a
way that the field
equation reduces to \be \fr{D}{D s}\fr{DA_i}{Ds}
+m^2A_i = 0\,\,, \ee 
the same
way as it happens in standard Cartesian coordinates
for Minkowskian 
spacetime.
Correspondingly, the field Lagrangian is $-\dot{\bf
A}^2 + m^2{\bf 
A}^2$,
where dot stands for the covariant derivative on $s$,
if and only if 
the curve
$\L$ is chosen properly. This Lagrangian should
contain one more term
responsible for the shape of $\L$, yielding the
geodesic equation when 
we
switch off the field. The form of this term is well
known:  $1/2 m\dot 
{\bf
x}^2$, thus our final Lagrangian is \be 2L = m \dot
{\bf x}^2 - \dot 
{\bf A}^2
+ m^2 {\bf A}^2 \,\,, \label{eq:lagrangian} \ee and
coupling between 
the field
and the shape of the curve $\L$ is incorporated in the
form of a 
covariant
derivative $\dot A$, which contains the product of
connection, velocity 
$\dot
x$ and the field.

The derivation of the conservation laws is more
convenient in 
orthonormal
frames. In what follows, $e^a_i$ will denote the
components of an 
orthonormal
1-form frame field, \be \t^a = e^a_i(x)dx^i\,\,, \ee
and $e_a^i$ the
components of its dual vector frame field, \be {\bf 
e}_a=e^i_a(x)\dd{x^i}\,\,.
\label{eq:e^i_a} \ee Here frame indices are always
$a,b,c,...$; 
coordinate
indices are $i,j,k,...$.  The metric tensor can be
expressed as \be
g=g_{ij}dx^i \otimes dx^j=\d_{ab}\t^a \otimes
\t^b\,\,.  \ee The 
connection
1-form for these frames may be introduced through the
first structure
equation:  \be d\t^a=\o_b^{\;\;a} \wedge
\t^b,\;\;\;\;\;\;\;\;\o_{ab}+\o_{ba}=0
\label{eq:cartan1} \ee and the
connection coefficients $\g_{abc}$ are that of the
expansion of this 
1-form in
the local frames $\left\{\t^a\right\}$:  \be
\o_b^{\;\;c}
=\g_{ab}^{\;\;\;c}\t^a\,\,.  \ee Thus, \be \dot A^a =
\fr{d A^a}{ds} +
\gamma_{bc}^{\;\;a}A^c\fr{dx^b}{ds}  \ee or $\dot
A^a$ is a covariant
derivative in orthonormal frame on a curve $x^i(s)$
with
$\gamma_{bc}^{\;\;\;a}$ being a spin connection.

\vspace*{1pt}\textlineskip \section{Equations of
Motion} 
\vspace*{-0.5pt}

Each generalized coordinate has its conjugate
generalized momentum:  
\be p_a
\equiv \pder{L}{\dot x^a} = m \eta_{ab} \dot x^b -
\eta_{db} \dot A^b 
A^c
\g_{ac}^{\;\;\;d} \,\,.  \ee and \be E_a \equiv
\pder{L}{\dot A^a} = 
-\dot A^b
\eta_{ab}\,\,, \label{eq:E_a} \ee Corresponding to
them the 
Euler-Lagrange
equations are \be \fr{d}{ds}p_a = \pder{L}{x^a} \,\,
\label{eq:p_a} \ee 
and
\be \fr{d}{ds}E_a = \pder{L}{A^a} \,\,. 
\label{eq:eulerforE_a} \ee Let 
us
first consider equations for the $E_a$
(\ref{eq:eulerforE_a}).  Right 
hand
side of this equation is:  \be \pder{L}{A^a} = -\dot
A^c \eta_{bc}
\g_{ca}^{\;\;\;b}\dot x^c + m^2 A^c\eta_{ac}\,.  \ee
And Euler-Lagrange
equations for $E_a$ will be:  \be \left(\fr{d}{ds}E_a
-E_a 
\g_{ca}^{\;\;\;b}
\dot x^c \right)- m^2 A^c\eta_{ac} = 0\,\,, \ee where
we used
Eq.~\ref{eq:E_a}.  Expression in the brackets is a
covariant derivative 
for
$E_a$, thus, we obtain:  \be \dot E_a - m^2
A^c\eta_{ac} = 0\,\,.  \ee 
Again
using Eq.~\ref{eq:E_a}, we finally obtain:  \be \ddot
A^b \eta_{ab} + 
m^2 A^b
\eta_{ab} = 0 \,\,, \ee which, as we can see, reduces
to Proca equation 
for
the four-vector field $A_{\mu}(x)$ \cite{QFTbook}: 
\be \left(\Box 
+m^2\right)
A_{\rho} =0\,\,.  \label{eq:proca} \ee

We describe spin of the particle in this model
directly by a tensor of 
spin
$S_{ab}$:  \be mS_{ab} = \fr{1}{2}\left(\dot A_b A_a -
\dot A_a A_b
\right)\,\,.  \label{eq:spin_def} \ee To obtain the
equation of motion for spin, we write 
\be m\fr{d S_{ab}}{ds} =
\fr{1}{2}\fr{d}{ds}\left[\dot 
A_a A_b -
\dot A_b A_a \right] \,\,.  \ee Due to (\ref{eq:proca})
all partial 
derivatives
of $A$ vanish and we are left with:  \berr & &
\hspace{-1.in}
m\fr{d S_{db}}{ds}
= \fr{1}{2} \dot x^a \left[\dot A_i A_j - \dot A_j A_i
\right] \dd{x^a}
\left(e^i_{d} e^j_{b}\right) = \no \\ & & \fr{1}{2}
\dot
x^c\left(\g_{cd}^{\;\;\;a} \left[\dot A_a A_b- \dot
A_b A_a\right] +
\g_{cb}^{\;\;\;a} \left[\dot A_d A_a - \dot A_a
A_d\right]\right)\,\,.  
\err
Here $e^i_a$ is the matrix introduced in the equation
(\ref{eq:e^i_a}) 
and
derivatives $\left(e^i_{b}\right)_{,a}$ are obtained
from the first 
structure
equation (\ref{eq:cartan1}).

Using our definition of a spin tensor
(Eq.~\ref{eq:spin_def}), we 
obtain:  \be
\fr{d S_{db}}{ds} = \dot x^c \left(\g_{cd}^{\;\;\;a}
S_{ab} +
\g_{cb}^{\;\;\;a} S_{ad} \right)\,\,, \ee or \be
\fr{DS_{db}}{Ds} = 0 
\,\,.
\label{eq:spin0} \ee Thus, spin is transported
parallel to itself along 
the
worldline.

To derive Euler-Lagrange equations for generalized
momentum, we return 
to the
coordinate basis.  In this case Lagrangian
(Eq.~\ref{eq:lagrangian}) 
will take
a form\footnote{It must be noted that while Eq.~2
represents a non-relativistic 
Lagrangian of the free mass point when we remove the
field part, 
here we use the relativistic form; in this case both
forms are 
actually equivalent. When we are considering
relativistic kinematics 
(free motion of a mass point, for example), the
trajectory 
will be a geodesic regardless of what form a
Lagrangian we take.}:  \be 2L=mg_{ij}\dot x^i \dot x^j
- g_{ij}\left(\fr{dA^i}{ds} +
\G^i_{kl} \dot x^k A^l\right) \left(\fr{dA^j}{ds} +
\G^j_{kl} \dot x^k
A^l\right) + m^2g_{ij}A^iA^j\,\,, \ee where
$\G^j_{kl}$ now are 
Christoffel
symbols.  Generalized momentum from here is:  \be p_i
\equiv 
\pder{L}{\dot
x^i} = m g_{ij}\dot x^j - \G^m_{il} A^l \dot A_m\,\,. 
\ee

In coordinate basis Eq.~\ref{eq:p_a} becomes:  \be
\fr{d}{ds}p_i =
\pder{L}{x^i} \,\,.  \ee

LHS of this equation is:  \be \fr{dp_i}{ds} = m
g_{ij}\fr{d \dot x^j 
}{ds} + m
\dot x^j \dot x^k \del_k g_{ij} - \dot x^j A^l \dot
A_m \del_j 
\G^m_{il} -
\G^m_{il}\left(\fr{dA^l}{ds}\dot A_m + A^l \fr{d\dot
A_m}{ds}\right) 
\;\;,
\label{eq:LHS} \ee where derivatives $\del_i$ are
defined as
$\del_i=\dd{x^i}$.  RHS is:  \be \pder{L}{x^i}=
\fr{1}{2} m \del_i 
g_{jk} \dot
x^j \dot x^k + \del_ig_{mn} \left(-\dot A^m \dot A^n +
m^2 A^m A^n 
\right) -
g_{mn} \dot x^k A^l \dot A^n \del_i \G^m_{kl} \,\,. 
\label{eq:RHS} \ee
Subtracting (\ref{eq:RHS}) from (\ref{eq:LHS}), we
obtain:  \berr & &
\hspace{-0.5in}m g_{ij}\fr{d \dot x^j }{ds} + m \dot
x^j \dot x^k 
\del_k
g_{ij} - \fr{1}{2}m \del_i g_{jk} \dot x^j \dot x^k -
\del_ig_{mn} m^2 
A^mA^n
- \G^m_{il} A^l \fr{d \dot A^m}{ds} + \no \\ & &
\hspace{-0.3in}\del_i 
g_{mn}
\dot A^m \dot A^n - \G^m_{il}\dot A_m\fr{dA^l}{ds} -
\dot x^j A^l \dot
A_m\del_j \G^m_{il} + g_{mn}\dot x^k A^l \dot A^n
\del_i \G^{m}_{kl} = 
0 \,\,.
\label{eq:E_L1} \err Using \be \del_i g_{mn}=\fr{1}{2}
\left(g_{kn}\G^k_{mi}
+g_{km}\G^k_{in}\right)\,\,, \label{eq:del_g} \ee and
some algebra, we 
obtain
from (\ref{eq:E_L1}) \berr & & \hspace
{-0.3in}mg_{ij}\fr{D \dot 
x^j}{Ds}
-\left( m^2 \G^k_{mi}A^m A_k + \G^m_{il} A^l \ddot A_m
\right) -
\G^m_{il}\G^k_{mj} \dot x^j A^l \dot A_k +
\G^n_{mi}\dot A^m \dot A_n 
-\no \\
& & \G^m_{il}\dot A^l \dot A_m + \dot x^j A^n \dot A_m
\G^m_{il}\G^l_{jn} -
\dot x^j A^l \dot A_m \del_j \G^m_{il} + \dot x^k A^l
\dot A_m
\del_i\G^m_{kl}=0\,\,.  \err We may notice that terms
in the brackets 
are zero
due to the Proca equation (\ref{eq:proca}) and terms
containing $\dot 
A\dot A$
cancel; the remaining terms \be g_{ij}m \fr{D \dot
x^j}{Ds}= \dot x^j 
A^l\dot
A_k \left(\del_j\G^k_{il} -\del_i\G^k_{jl} +
\G^k_{im}\G^m_{jl} -
\G^k_{mj}\G^m_{il} \right)\,\,.  \ee With the
definitions of the 
curvature
tensor and spin tensor (\ref{eq:spin_def}), the
equation takes the 
evidently
covariant form:  \be g_{ij}\fr{D\dot x^j}{Ds} =
R^{k}_{\;\;\;\;jil} 
\dot x^j
S_k^{\;\;l} \,\,.  \label{eq:pap1} \ee This equation
coincides with the
Papapetrou equation \cite{Papapetrou}. It
must be pointed out that in case of a zero spin this
equation becomes a
geodesic.  If we reparametrise the curve with some new
parameter $\l$ 
in such
a way that \be g_{ij}\fr{dx^i}{d\l}\fr{dx^j}{d\l}
=0\,\,, \ee we can 
rewrite
Eq.~\ref{eq:pap1} as \be g_{ij}\fr{D\dot x^j}{D\l} = 
R^{k}_{\;\;\;\;jil} \dot
x^j S_k^{\;\;l} \,\,. \ee This equation is valid for
the massless 
particles
with spin as well, since mass does not enter the
equation explicitly.

\vspace*{1pt}\textlineskip \section{Acknowledgments}
\vspace*{-0.5pt}

ZT wishes to thank ICTP, Trieste for a travel grant
NET-53.  MS is 
supported
by an ICCR scholarship (Indo-Russian Exchange
programme) and authors 
thank the
hospitality of IUCAA, Pune, where this work has been
carried out.

\vspace*{1pt}\textlineskip  \end{document}